\documentclass[aps,prd,twocolumn,nofootinbib,showpacs]{revtex4}
\usepackage{graphicx}
\usepackage{mathrsfs}
\usepackage{amsmath}
\usepackage{textcomp}

\usepackage[colorlinks=true, linkcolor=blue, citecolor=blue, urlcolor=blue]{hyperref}

\begin{document}

\title{Novel analysis for the energy-energy correlation \\ in electron-positron annihilation in the perturbative domain}

\author{Zhu-Yu Ren$^1$}

\author{Sheng-Quan Wang$^1$}
\email[email:]{sqwang@cqu.edu.cn}

\author{Jian-Ming Shen$^{2}$}
\email[email:]{shenjm@hnu.edu.cn}

\author{Xing-Gang Wu$^3$}
\email[email:]{wuxg@cqu.edu.cn}

\author{Leonardo Di Giustino$^{4,5}$}
\email[email:]{leonardo.digiustino@uninsubria.it}

\author{Philip G. Ratcliffe$^{4,5}$}
\email[email:]{philip.ratcliffe@uninsubria.it}

\author{Stanley J. Brodsky$^6$}
\email[email:]{sjbth@slac.stanford.edu}

\address{$^1$Department of Physics, Guizhou Minzu University, Guiyang 550025, P.R. China}
\address{$^2$School of Physics and Electronics, Hunan University, Changsha 410082, P.R. China}
\address{$^3$Department of Physics, Chongqing University, Chongqing 401331, P.R. China}
\address{$^4$Department of Science and High Technology, University of Insubria, via Valleggio 11, I-22100, Como, Italy}
\address{$^5$INFN, Sezione di Milano-Bicocca, 20126 Milano, Italy}
\address{$^6$SLAC National Accelerator Laboratory, Stanford University, Stanford, California 94039, USA}

\date{\today}

\begin{abstract}
The energy-energy correlation (EEC) in electron-positron annihilation plays a crucial role in precision tests of quantum chromodynamics (QCD) and measurements of the QCD coupling constant. In this paper, we provide a novel analysis for the EEC by using the Principle of Maximum Conformality (PMC), a systematic method for eliminating renormalization scheme-and-scale ambiguities. The PMC scales are determined by resumming the non-conformal $\beta$-terms that govern the behavior of the QCD running coupling via the renormalization group equation, and reflect the virtuality of the propagating gluons in QCD. It is noteworthy that the resulting PMC scale varies dynamically with the EEC's angular distribution, reflecting the expected scale's physical behavior. Moreover, due to the reabsorption of all $\beta$-terms, including also those related to the divergent renormalon terms such as $n!\beta^n_0\alpha^n_s$, in the pQCD series, the behavior of the QCD perturbative coefficient using PMC, differs entirely from that of the conventional coefficient. Consequently, the PMC predicted EEC distribution agrees well with the experimental data in the perturbative domain.
\end{abstract}

\maketitle

Event shape variables in electron-positron annihilation are a fundamental tool for precision phenomenological studies of quantum chromodynamics (QCD) and for high precision measurements of the QCD coupling constant. Among the various event shape variables, the energy-energy correlation (EEC) function~\cite{Basham:1978bw,Basham:1977iq,Basham:1978zq,Basham:1979gh} has attracted considerable interest from both the experimental and theoretical communities over the years (see e.g.~\cite{Moult:2025nhu} for a review). Defined as a quantity weighted by the product of the energies of the detected particles, the EEC is measured using two detectors positioned at a fixed angular separation. The EEC is defined as:
\begin{eqnarray}
\frac{d\sum}{d\cos{\chi}}=\sum_{ij}\int\frac{E_{i}E_{j}}{Q^2}\delta(\cos{\chi-\cos{\theta_{ij}}})d\sigma,
\label{eq1}
\end{eqnarray}
where, $Q$ is the total center-of-mass energy, $E_i$ and $E_j$ are the particle energies,  $\theta_{ij}$ is the angle between the two detected particles.

Numerous experimental groups have achieved high-precision measurements for the EEC~\cite{OPAL:1993pnw,SLD:1994idb,L3:1992btq,OPAL:1990reb,L3:1991qlf,SLD:1994yoe,PLUTO:1985yzc,PLUTO:1979vfu,CELLO:1982rca,JADE:1984taa,TASSO:1987mcs,TOPAZ:1989yod,ALEPH:1990vew,DELPHI:1990sof,Fernandez:1984db,Wood:1987uf}. To explain the experimental data, it is essential to have equally high-precision theoretical predictions. Fixed-order QCD calculations of the EEC have been thoroughly investigated in the existing literature~\cite{Schneider:1983iu,Falck:1988gb,Glover:1994vz,Kramer:1996qr,Ali:1982ub,Ali:1984gzn,Richards:1982te,Richards:1983sr,Dixon:2018qgp,Tulipant:2017ybb,DelDuca:2016csb,DelDuca:2016ily}. In recent years, significant advances have been achieved in the study of the back-to-back regions~\cite{Jaarsma:2025tck,Tulipant:2017ybb,Collins:1981uk,Collins:1981va,Collins:1985xx,Collins:1985kw,Collins:1981zc,Kodaira:1981nh,Kodaira:1982az,deFlorian:2004mp,Kardos:2018kqj,Korchemsky:2019nzm,Aglietti:2024xwv,Moult:2018jzp,Ebert:2020sfi,Duhr:2022yyp,Kang:2024dja,Kang:2023big,Chao:1982wb,Soper:1982wc}. Separately, the resummation of large infrared logarithms in the collinear regions has also been addressed in Refs.~\cite{Jaarsma:2025tck,Korchemsky:2019nzm,Dixon:2019uzg,Guo:2025zwb}, while non-perturbative effects and their implications have been explored in Refs.~\cite{Jaarsma:2025tck,Herrmann:2025fqy,Nason:1995np,Korchemsky:1999kt,Belitsky:2001ij,Dokshitzer:1999sh,Dokshitzer:1998pt,Dokshitzer:1997iz,Abbate:2010xh,Lee:2006nr,Salam:2001bd,Mateu:2012nk,Schindler:2023cww,Lee:2024esz,Cuerpo:2025zde,Liu:2024lxy}.

Currently, the main obstacle for achieving highly precise EEC predictions and the QCD coupling determination from the EEC process is not a lack of precise experimental data, but the ambiguity of theoretical predictions. According to the conventional scale-setting procedure, in order to eliminate the large logarithmic terms $\ln({\mu_r/Q})$ from the pQCD prediction, the renormalization scale $\mu_r$ for the EEC distribution is simply set to the center-of-mass energy $Q$. The corresponding theoretical uncertainty is then estimated by varying the scale within an arbitrary range, typically $\mu_r\in [Q/2,2Q]$. It is well-known that conventional theoretical predictions for the EEC are plagued by the renormalization scale uncertainty, and even at NNLO the conventional predictions do not fit the experimental data. Despite dedicated efforts to mitigate these issues, the renormalization scale uncertainty remains the dominant factor limiting the accuracy of theoretical predictions.
By using the renormalization group invariance argument, one would expect that variations in the renormalization scale should be sensitive only to non-conformal terms and not to conformal ones. In addition, the range of the scale variation within which reliable predictions can be obtained remains unspecified. The conventional practice is strictly scale and scheme dependent, being entangled with the renormalization scale and scheme ambiguities of the perturbative QCD series, and thus the predictions obtained are significantly affected by scale and scheme uncertainties.

The nature of the EEC distribution corrections in the collinear and back-to-back regions is basically non-perturbative.  Physically, the renormalization scale, which reflects the subprocess virtuality, should become soft in these two regions. This effect is simply ignored by the conventional approach, the renormalization scale being set to $\mu_r=Q$ over the entire angular spectrum of the process, which leads to a violation of the correct physical behavior in the regions affected by non-perturbative effects and thus to unreliable predictions. It is therefore highly desirable to develop a method that eliminates the renormalization scale ambiguity, thereby improving the accuracy of theoretical predictions.

The Principle of Maximum Conformality (PMC)~\cite{Brodsky:2011ta,Brodsky:2012rj,Brodsky:2011ig,Mojaza:2012mf,Brodsky:2013vpa} provides a systematic method for eliminating ambiguities associated with the renormalization scheme and scale in perturbative predictions. The PMC is the  principle underlying the well-known Brodsky-Lepage-Mackenzie (BLM) method~\cite{Brodsky:1982gc}.
The BLM and PMC methods match at NLO and in the Abelian limit reduce to the well-known Gell-Mann-Low method~\cite{Gell-Mann:1954yli}. The PMC scales are unambiguously determined by absorbing the $\beta$ terms that govern the behavior of the running coupling via the Renormalization Group Equations (RGE), and thus the PMC scale reflects the virtuality of gluon propagators at a given order and consequently sets the effective number $n_f$ of active flavors. The resulting fixed-scale perturbative series is identical to the corresponding ``conformal" series with $\beta\equiv0$. Since all $\beta$ terms are cancelled in the PMC pQCD series, the renormalon divergence is eliminated, this generally leads to a PMC pQCD series having improved convergence.

The PMC method has been successfully applied, eliminating scale uncertainties, to the six typical event-shape variables in electron-positron annihilation~\cite{Wang:2019ljl,Wang:2019isi,Wang:2021tak,Ren:2025uns,DiGiustino:2020fbk,DiGiustino:2021nep,DiGiustino:2024zss}. The PMC provides a notable way to determine the QCD running coupling $\alpha_s(Q)$ from the event-shape distributions measured at a single energy, i.e. at the energy of the process $Q$~\cite{Wang:2019isi,DiGiustino:2024zss}. It has been found that the PMC can be adopted to consistently describe both total and differential observables~\cite{Huang:2024zni}.

The renormalization scale in Quantum Electrodynamics (QED) is set unambiguously by using the Gell-Mann-Low method~\cite{Gell-Mann:1954yli} where the renormalization scale is given by the virtuality of the exchanged photon propagator. It is noted that the PMC scale in QED is identical to the QCD PMC scale in the physical $V$ scheme, which we shall describe shortly. A robust pQCD prediction for event-shape variables can be achieved by using the PMC method together with the use of the physical $V$ scheme~\cite{Ren:2025uns}. In this paper, we use this strategy, presenting a detailed analysis of the EEC by using the PMC method and adopting the physical $V$ scheme.

At NLO, the perturbative expansion of the EEC distribution in the $\overline{\rm MS}$ scheme and with the initial renormalization scale set to $\mu_r=\sqrt{s}=Q$, is written as:
\begin{eqnarray}
\frac{1}{\sigma_{0}}\frac{d\sum}{d\cos{\chi}}=A(z)\,a_s(Q)+B(z)\,a^2_s(Q)+{\cal O}(a^3_s),
\label{eq3}
\end{eqnarray}
where $a_s(Q)=\alpha_s(Q)/2\pi$, $\sigma_0$ denotes the Born cross section for the process $e^+e^-\to q\bar q$, $A(z)$ and $B(z)$ are the LO and NLO coefficients respectively, which are analytically reported in Ref.~\cite{Dixon:2018qgp}. Here, $z$ is a single variable defined explicitly by $z=(1-\cos{\chi})/2$. The distribution of the EEC exhibits two well-defined peaks at $\chi=0$ ($z=0$) and $\chi=\pi$ ($z=1$). These two peaks are in fact a phenomenological effect of the infrared(IR)-large-logarithms resulting from the emission of soft and collinear partons. In order to improve the description of the EEC distribution given by the pQCD expansion in these two regions, a resummation technique would be required. In contrast, at intermediate angles, higher-order corrections tend to broaden the distribution.

The EEC distribution is typically normalized to the total hadronic cross section:
\begin{eqnarray}
\sigma_{t}=\sigma_0\left(1+\tfrac{3}{2}C_F\,a_s(Q)+{\cal O}(a^3_s)\right),
\end{eqnarray}
the color factor $C_F=\frac{4}{3}$, thus, Eq.(\ref{eq3}) becomes:
\begin{eqnarray}
\frac{1}{\sigma_{t}}\frac{d\sum}{d\cos{\chi}}=\bar A(z)\,a_s(Q)+\bar B(z)\,a^2_s(Q)+{\cal O}(a^3_s).
\label{eq5}
\end{eqnarray}
The normalized coefficients $\bar A(z)=A(z)$, $\bar B(z)=B(z)-\frac{3}{2}C_FA(z)$. The dependence on the general renormalization scale $\mu_r$, i.e., $\bar A(z,\mu_r)$, $\bar B(z,\mu_r)$ for the perturbative coefficients, is obtained via the renormalization group equation. At NLO, the coefficient $\bar B(z,\mu_r)$ can be further separated into two parts, one dependent on the number of active quark flavours, $n_f$, and one independent of $n_f$, i.e.,
\begin{eqnarray}
\bar B(z,\mu_r)=\bar B_{in}(z,\mu_r)+\bar B_{n_f}(z,\mu_r)\cdot n_f.
\label{eq6}
\end{eqnarray}

Before implementing the PMC method in the EEC distribution, we first transform the perturbative series from the $\overline{\rm MS}$ scheme to the $V$ scheme. In contrast to the non-physical $\overline{\rm MS}$ scheme, the $V$ scheme defined by the QCD static potential between two heavy quarks~\cite{Appelquist:1977tw,Fischler:1977yf,Schroder:1998vy}, is a physical renormalization scheme. It is particularly suitable to obtain a more physical definition of the strong coupling defined by the effective-charge. Moreover, the PMC scale in QED is the same as the QCD PMC scale in the physical $V$ scheme~\cite{Wang:2020ckr}. A robust pQCD prediction for the event shape variables is achieved by using the PMC method together with the use of the $V$ scheme~\cite{Ren:2025uns}. Here, we apply this approach to the EEC distribution. The PMC predictions are independent of the scheme, which is guaranteed by the conformal PMC series and are clearly presented in the form of ``commensurate scale relations" (CSR)~\cite{Brodsky:1994eh}.

The relation of the coupling constant between the $\overline{\rm MS}$ scheme and the $V$ scheme is given by
\begin{eqnarray}
a_s(Q) = \sum_{i=1}^{\infty} r^{V}_{i} a_s^{V,i}(Q),
\label{eq7}
\end{eqnarray}
the first two perturbative coefficients are~\cite{Anzai:2009tm,Smirnov:2009fh,Smirnov:2008pn,Kataev:2015yha,Kataev:2023sru}
\begin{eqnarray}
r^{V}_1=1, \,\,\,\,r^{V}_2=-\tfrac{31}{18}C_A+\tfrac{10}{9}T_F\cdot n_f,
\label{eq8}
\end{eqnarray}
using the Eq.(\ref{eq7}), we can transform the EEC into the physical $V$ scheme,
\begin{align}
\frac{1}{\sigma_{t}}\frac{d\sum}{d\cos{\chi}}=&\bar A(z,\mu_r)^{V}a_s^V(\mu_r)+\bar B(z,\mu_r)^{V}a^{V,2}_s(\mu_r)\nonumber \\&+{\cal O}(a^{V,3}_{s})
\label{eq9}
\end{align}
where $\bar A(z,\mu_r)^{V}=\bar A(z)$. Similarly to Eq.(\ref{eq6}), the coefficient $\bar B(z,\mu_r)^{V}$ in the $V$ scheme is
\begin{eqnarray}
\bar B(z,\mu_r)^V=\bar B(z,\mu_r)^V_{in}+\bar B(z,\mu_r)^{V}_{n_f}\cdot n_f.
\end{eqnarray}
At NLO level, the number of active quark flavours $n_f$ is related to the $\beta_0$ term by $\beta_0=11-2/3\,n_f$. After absorbing the $\beta_0$ term using the PMC, we obtain the following final conformal form:
\begin{align}
\frac{1}{\sigma_{t}}\frac{d\sum}{d\cos{\chi}}=&\bar A(z,\mu_r)^Va^V_s(Q_\star)+\bar B(z,\mu_r)^V_{con}a^{V,2}_s(Q_\star)\nonumber\\& +{\cal O}(a^{V,3}_s),
\label{eq11}
\end{align}
where the conformal coefficient can be written as
\begin{eqnarray}
\bar B(z,\mu_r)^V_{con}=\frac{33}{2}\bar B(z,\mu_r)^V_{n_f}+\bar B(z,\mu_r)^V_{in}
\end{eqnarray}
and the PMC scale is given by:
\begin{eqnarray}
\ln\frac{Q_\star}{\mu_r}=\frac{3\bar B(z,\mu_r)^V_{n_f}}{2\bar A(z,\mu_r)^V}.
 \end{eqnarray}

After applying the PMC, only the conformal terms remain in the pQCD series. Consequently, the resulting perturbative series is independent of the choice of the renormalization scale $\mu_r$, and the PMC scale itself is also scale-independent. Thus, in Eq.(\ref{eq11})  the renormalization scale ambiguity for the EEC distribution is eliminated.

Numerical calculations are performed adopting the two-loop QCD coupling constant, of which the asymptotic scale, $\Lambda^{\overline{MS}}_{QCD}$, is determined by using the world-average value $\alpha_s(M_z)=0.1180$~\cite{ParticleDataGroup:2024cfk}. The asymptotic scale in the $V$ scheme is obtained by the shift of scale: $\Lambda^V_{QCD}=\Lambda^{\overline{MS}}_{QCD}\exp\left[\left(\frac{31}{6} - \frac{5}{9} n_f \right)/\beta_0\right]$.

\begin{figure} [htb]
\centering
\includegraphics[width=0.40\textwidth]{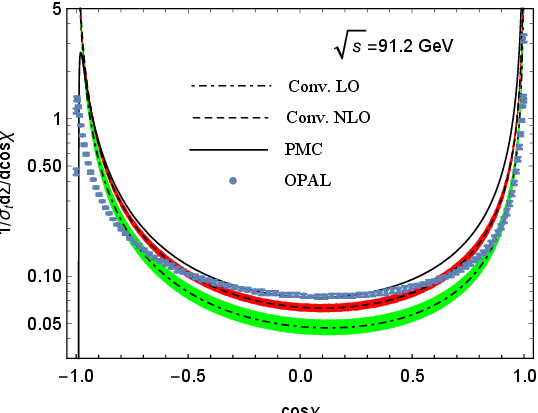}
\caption{The EEC differential distributions using the conventional and the PMC scale setting at $\sqrt{s}=91.2$ GeV. The bands for theoretical predictions are obtained by varying $\mu_r\in [\sqrt{s}/2,2\sqrt{s}]$. The PMC prediction is independent of the scale $\mu_r$.}
\label{Conv}
\end{figure}

We present the differential distribution for the EEC function using conventional and PMC scale settings at the center-of-mass energy $\sqrt{s}=91.2~\text{GeV}$ in Fig.~\ref{Conv}. For the case of conventional scale setting, the renormalization scale $\mu_r$ is simply set to $\mu_r=\sqrt{s}$, and the theoretical uncertainties are estimated by varying the scale within the range $\mu_r\in [\sqrt{s}/2,2\sqrt{s}]$. As clearly illustrated in Fig.~\ref{Conv}, conventional predictions up to NLO are plagued by a significant renormalization scale uncertainty. Even up to NNLO in QCD corrections\cite{Tulipant:2017ybb} in the intermediate-angle perturbative region, the calculations at NLO do not overlap with the LO predictions and those at NNLO do not overlap with the NLO predictions either. This is mainly due to the fact that the variation of the scale is only sensitive to non-conformal $\beta$-terms, it follows that the estimate of unknown higher-order QCD contributions by varying the scale $\mu_r\in[\sqrt{s}/2,2\sqrt{s}]$ is unreliable. Furthermore, the conventional perturbative series for the EEC distribution shows a slowly convergent behavior also owing to the presence of renormalon terms.

The nature of the corrections to the EEC distribution in the collinear ($\chi=0$) and back-to-back ($\chi=\pi$) regions is essentially non-perturbative. Physically, the renormalization scale should become very soft in these two regions.
By simply setting the renormalization scale to $\mu_r=\sqrt{s}$, one cannot unambiguously separate perturbative and non-perturbative physics, leading inevitably to an incomplete description and to unreliable predictions especially in estimating the errors. Moreover, even at NNLO, conventional predictions do not fit the experimental data.

After applying the PMC, the EEC distribution is significantly increased and its behavior is different from that of the conventional scale setting. We note that in the back-to-back region, the experimental measurements clearly show that the EEC first increases and then decreases as the angle increases. The conventional predictions do not display this behavior. In contrast, although there are some deviations from the experimental measurements, the PMC predictions have the same overall behavior, i.e. first increasing and then decreasing when approaching the back-to-back region, as clearly illustrated in Fig.~\ref{Conv}.

Currently, pQCD predictions for the EEC are known up to NNLO. However, how to fully distinguish the conformal and non-conformal contributions at NNLO for the EEC is less straightforward. We note that the PMC scale, $Q_\star$, results from a perturbative expansion in $\alpha_s$; it can be determined up to leading-log (LL) and next-to-leading-log (NLL) accuracy using the non-conformal contributions at NLO and NNLO, respectively. The PMC scales $Q_\star$ for all event shapes show fast pQCD convergence, and inclusion of the NNLO contributions only slightly changes the PMC scales $Q_\star$ at LL~\cite{Wang:2019ljl,Wang:2021tak}. In addition, the $n_f$-dependent terms which are related to the $\beta_0$ terms, are clearly non-conformal contributions at NLO. Thus, we provide a NLO PMC analysis for the EEC.

\begin{figure} [htb]
\centering
\includegraphics[width=0.40\textwidth]{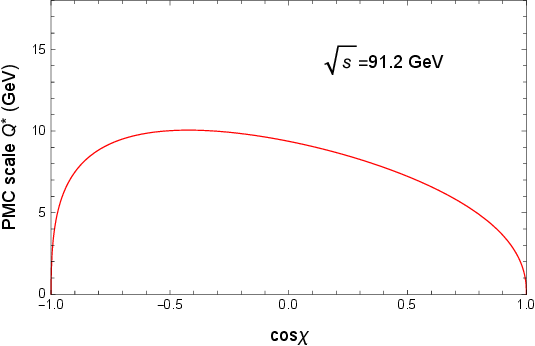}
\caption{The PMC scale for the EEC differential distribution at a center-of-mass energy $\sqrt{s}=91.2~\text{GeV}$ . }
\label{pmcscale}
\end{figure}

In Fig.~\ref{pmcscale} we present the PMC scale for the EEC at $\sqrt{s}=91.2$ GeV. The PMC scale is substantially smaller than the center-of-mass energy $\sqrt{s}$ and increases as $\sqrt{s}$ increases. Remarkably, unlike the conventional method,  the PMC scale is not a single-valued function but changes dynamically with the angular distribution of the EEC, fixing the effective number of active flavors $n_f$. Moreover, in the intermediate perturbative region, the PMC scale is large, while the PMC scale is very soft and gives rise to the zero-limit in the collinear ($\chi=0$) and back-to-back ($\chi=\pi$) regions reflecting the correct kinematic constraint at the border.

The PMC method enables interpolation across the full angular range of the EEC, thereby identifying the regime where non-perturbative QCD effects become significant. In contrast, conventional scale-setting provides an incomplete description and inadequate control over the physical range of the EEC and the emergence of detrimental non-perturbative contributions. Specifically, in the collinear ($\chi=0$) and back-to-back ($\chi=\pi$) regions, the PMC scale approaches a soft limit, signaling the presence of non-perturbative effects, such as large infrared logarithms related to soft-and-collinear divergences. Thus, in order to improve the perturbative results and the description of the physical phenomena, a resummation technique would be required. This is consistent with soft-collinear effective theories (such as SCET)~\cite{Bauer:2000ew,Bauer:2000yr,Bauer:2001yt,Beneke:2002ph,Aglietti:2007bp,DiGiustino:2011jn,Aglietti:2025jdj}, which predict the presence of soft scales in the collinear and back-to-back regions.

\begin{figure} [htb]
\centering
\includegraphics[width=0.23\textwidth]{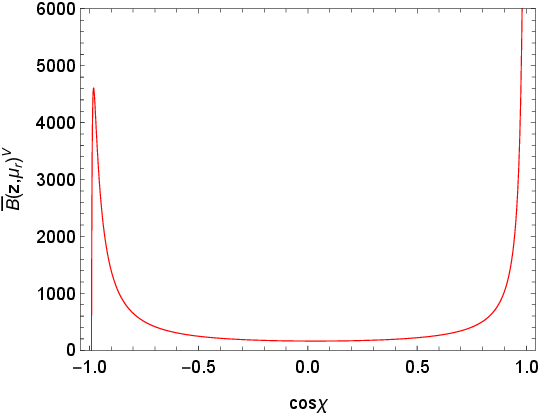}
\includegraphics[width=0.23\textwidth]{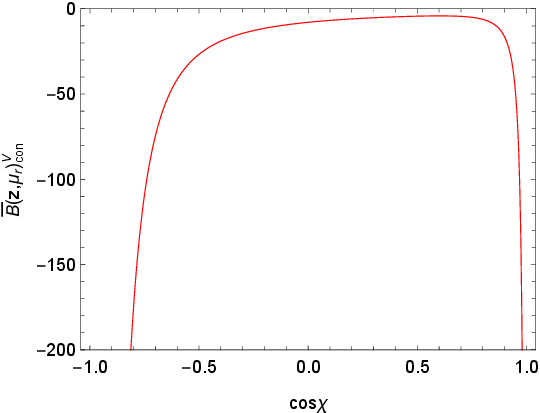}
\caption{The NLO perturbative coefficients using the conventional (left) and the PMC (right) scale setting. }
\label{coefficient}
\end{figure}

In addition, also the behavior of the PMC-NLO perturbative coefficient differs significantly from the corresponding conventional coefficient; this is an effect of the $\beta$-term (i.e. the divergent renormalon term) cancellation in the pQCD series. In Fig.~\ref{coefficient}, we present the comparison between the conventional and the PMC NLO perturbative coefficient. From Fig.~\ref{coefficient} we notice a general opposite behavior for the NLO coefficient with respect to the scale-setting method, especially in the central perturbative region and in the back-to-back region.

\begin{figure} [htb]
\centering
\includegraphics[width=0.23\textwidth]{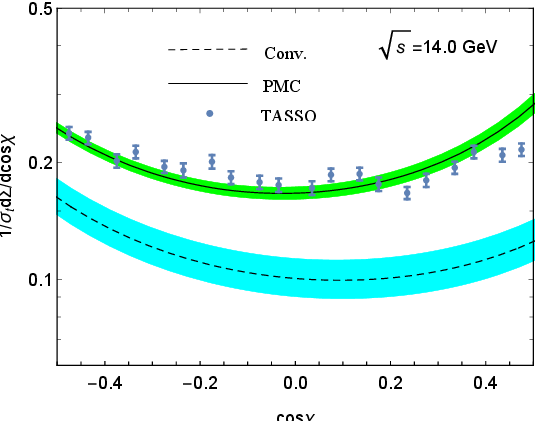}    
\includegraphics[width=0.23\textwidth]{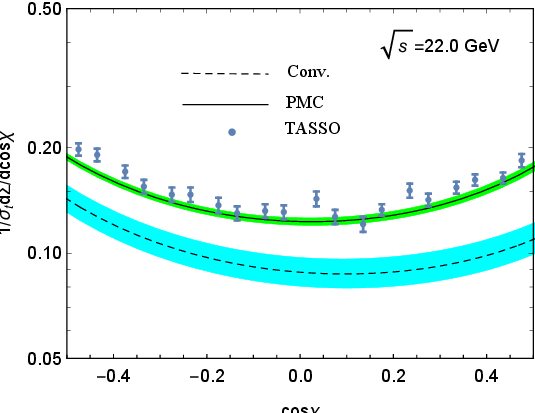}    
\includegraphics[width=0.23\textwidth]{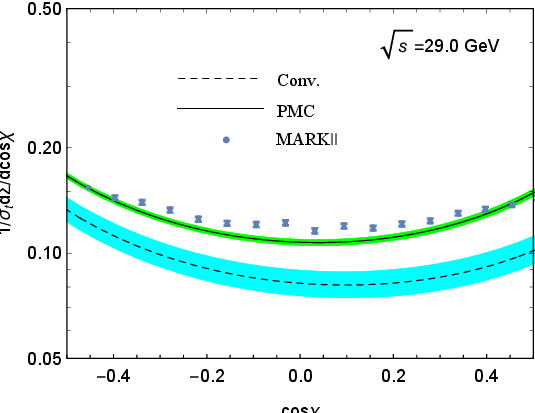}    
\includegraphics[width=0.23\textwidth]{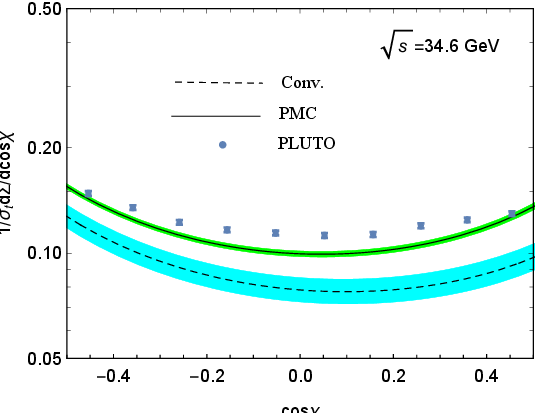} 
\includegraphics[width=0.23\textwidth]{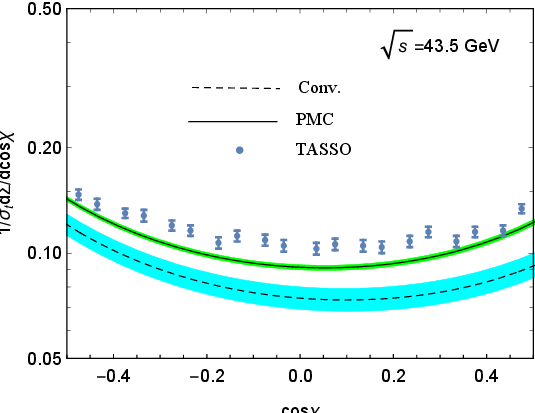} 
\includegraphics[width=0.23\textwidth]{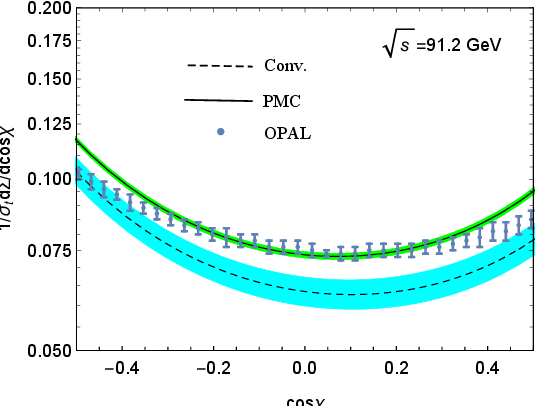}        
\caption{Results for the EEC distribution for $\sqrt{s}=14,~22,~29,~34.6,~43.5,91.2$ GeV in the intermediate perturbative region using conventional and PMC scale-settings. Error bands are the squared averages of the renormalization scale $\mu_r\in [\sqrt{s}/2,2\sqrt{s}]$ and the coupling constant $\Delta \alpha_s(M_Z)=\pm0.0009$~\cite{ParticleDataGroup:2024cfk}. The experimental data are taken from TASSO~\cite{TASSO:1987mcs}, MARKII~\cite{Fernandez:1984db}, PLUTO~\cite{PLUTO:1985yzc} and  OPAL~\cite{OPAL:1993pnw} .}
\label{pmc}
\end{figure}

Due to the scale and the NLO perturbative coefficient derived from the PMC scale setting being considerably different from the results obtained by the conventional scale setting, the resulting PMC predictions are substantially different from the conventional results. In Fig.~\ref{pmc}, we present the EEC distributions in the intermediate perturbative region at $\sqrt{s}=14,~22,~29,~34.6,~43.5,91.2$ GeV, along with the corresponding experimental data taken from TASSO~\cite{TASSO:1987mcs}, MARKII~\cite{Fernandez:1984db}, PLUTO~\cite{PLUTO:1985yzc}, OPAL~\cite{OPAL:1993pnw} collaborations and including a comparison between the conventional and PMC scale-setting results.
As shown in Fig.~\ref{pmc}, in the intermediate region, the PMC significantly improves the precision of the predictions eliminating renormalization-scale uncertainties and the results are in good agreement with experimental data, while conventional predictions are significantly affected by large scale uncertainties and underestimate experimental data. The lower-energy data, below the $Z^0$ peak for the EEC, are widely used to determine the QCD running coupling $\alpha_s$. Fig.~\ref{pmc} also shows that the PMC predictions have been substantially increased, especially for small $\sqrt{s}$. The PMC provides a rigorous, comprehensive description of the EEC over the wide range of center-of-mass energies $\sqrt{s}$ without artificial parameters.

Conclusions: we present a novel analysis of the EEC function in electron-positron annihilation using the PMC scale-setting method. In the conventional approach, the renormalization scale is simply set to the center-of-mass energy $Q$, with its uncertainty estimated by varying the scale in an arbitrary range. This procedure introduces inherent scheme and scale dependence in theoretical predictions, violating the renormalization group invariance principle. It also obscures the exact QCD correction terms at each order and hinders assessment of the intrinsic convergence of the perturbative QCD series. Moreover, the conventional method fails to capture the correct physical behavior of the scale near the collinear and back-to-back regions, inevitably leading to unreliable predictions for the EEC distribution.
By applying the PMC, renormalization scheme and scale ambiguities are eliminated. The PMC scale, determined by the RGE, is not a single-valued function but dynamically evolves with the EEC angular distribution over the full angular range, yielding a physically consistent description of the renormalization scale behavior. The PMC conformal coefficients exhibit distinct behavior compared to the conventional coefficients. The resulting PMC predictions agree well with experimental data in the intermediate angular region and show consistent behavior with data in the back-to-back regions. These results provide valuable input for future collider programs. We anticipate further applications of the PMC to other fundamental processes involving the EEC event-shape variable, including electron-proton, proton-antiproton, and proton-proton collisions.

\hspace{1cm}

{\bf Acknowledgements}: This work was supported in part by the Natural Science Foundation of China under Grant No.12265011, No.12575080 and No.12547101; by the Project of Guizhou Provincial Department of Science and Technology under Grant No.YQK[2023]016, No.ZK[2023]141 and No.CXTD[2025]030; by the Hunan Provincial Natural Science Foundation with Grant No.2024JJ3004, YueLuShan Center for Industrial Innovation (2024YCII0118).

\end{document}